# Estimation of Stability Regions of Droop Control Slopes for MMC-based MTDC Systems

Yuntao Zou, *Student Member, IEEE*, Lei Zhang, *Student Member, IEEE*, Jiangchao Qin, *Senior Member, IEEE*

*Abstract*—This paper proposes a computational method to efficiently and quickly estimate stability regions of droop control slopes for modular multilevel converter (MMC)-based multi-terminal dc (MTDC) systems. The proposed method is based on a general small-signal model consisting of a dc grid with arbitrary topology and MMCs with $dq$ controllers. The general small-signal model developed by a systematic way can be used for small-disturbance stability analysis. To verify the developed small-signal model, a comparison between the developed model calculated in MATLAB and the detailed switching model simulated in PSCAD/EMTDC is conducted, which demonstrates the accuracy of the developed small-signal model. Based on the eigenvalues sensitivity and the Taylor Series of eigenvalues, a set of inequality constraints are derived and used to efficiently estimate the stability regions of all coupled slopes of the droop characteristics. It is helpful for efficiently designing and adjusting the droop controller parameters for the MMC-MTDC systems. The effectiveness of the proposed method is demonstrated by the several examinations including the supremum test and the stability region sketch on accuracy and feasibility.

*Index Terms*—Modular multilevel converter (MMC), multi-terminal dc system (MTDC), stability analysis, stability region, eigenvalue sensitivity.

## I. Introduction

NOWADAYS, with the integration of renewable energy and the increased demands of bulk power transmission, the modular multilevel converter (MMC)-based multi-terminal dc (MTDC) system with droop control and intricate dc network [1]–[6] develops into a complicated nonlinear system, which increases the difficulty of stability analysis. Furthermore, the dynamic feedback loops of the converter stations of the MMC-MTDC system significantly affect small-signal stability.

Droop control is a common method for controlling the MTDC systems [7]. To ensure stable operation, the stability regions of system parameters need to be estimated for properly control design and safe operations. Many efforts have been done to explore the stability regions of system parameters. Reference [4] derives a simple constraint of the droop slope based on the existing condition of equilibrium points, which only reflects a very limited region. Bifurcation theory is an advanced method for searching the parameter stability region by exploring the boundary of Hopf bifurcation and saddle-node bifurcation [8], [9]. However, the computational burden of this method is quite heavy, especially for high-order systems, e.g., the MMC-MTDC systems. Reference [8] studies the parameter stability region of droop-controlled ac microgrid and improves the computing efficiency by applying the reduced-order model. Reference [9] selects a common bus as the topology of dc gird and simplifies the whole grid into a second-order system. Although the reduced-order models may improve the computational efficiency, these models are inaccurate for estimating stability region.

The parametric approximation is also employed to approximate the boundary of parameter stability region. Kernel Ridge Regression and Galerkin method are utilized in [5] and [10], respectively, for the stability region of bus voltages and active power. However, the heavy computational burden still cannot be avoided.

In this paper, a set of second-order inequality constraints are proposed based on the eigenvalue sensitivity analysis and the Taylor series of eigenvalues. These inequality constraints can efficiently and precisely estimate the stability regions of multiple droop slopes for the MMC-MTDC system with an analytic expression.

To ensure the accuracy of estimating stability region based on the proposed method, the arm switching function (ASF) for modeling the MMC is employed [2], [11]–[13]. For modeling of dc network, the existing literatures focus on the specific toplogies, e.g., the common bus [4], [6], two-terminal system [14], and four-terminal grid [2], [13]. To model the dc network with the arbitrary topology, a general method is presented based on the edge-node incidence matrix [15], which covers dc load buses and interconnecting buses. Moreover, a systematic way is presented to develop a general small-signal model of the MMC-MTDC system for stability analysis and region identification.

## II. Modeling of the MMC-MTDC System

In this section, the mathematical model of the MMC-MTDC system is developed. It consists of multiple MMC stations with the droop and $dq$ controllers, and dc network with the arbitrary topology.

*Remark*: if there is a sequence of variable $X$ with subscript $y$ and superscript $z$ shown as $X_{y1}^z, X_{y2}^z, \ldots, X_{yi}^z$ where $i$ is the integer, a notation for the vector and diagonal matrix comprised by them will be respectively expressed as

$$X_{\overrightarrow{[y]}}^z = \left[X_{y1}^z, X_{y2}^z, \ldots, X_{yi}^z\right]^T.$$
$$X_{\Lambda[y]}^z = diag\left(\left[X_{y1}^z, X_{y2}^z, \ldots, X_{yi}^z\right]\right)^T.$$

### A. Modeling of the MMC

The structure of an MMC station is shown in Fig. 1, where $L_s$ and $C_g$ are the smoothing reactor and grounding capacitor at the dc side, respectively. $v_{dc}$ is the dc-side voltage of the MMC and $v_{dc\_bus}$ is the dc grid voltage. To model the MMC, the ASF model of the MMC is employed and shown in Fig. 2 for phase $j$, where $j$ = a, b, and c. $N$ is the number of sub-modules in each arm and $C_{SM}$ is the sub-module (SM) capacitance. $L_{arm}$ and $R_{arm}$ are the arm inductance and resistance, respectively. $L_0$ and $R_0$ are the inductance and resistance of point of common coupling (PCC), respectively. The dynamics of the arm currents and voltages are, respectively, described by

$$v_{dc} - S_{jp}v_{jp} - R_{arm}i_{jp} - L_{arm}\frac{di_{jp}}{dt} = v_j^{ac},$$
$$-v_{dc} + S_{jn}v_{jn} + R_{arm}i_{jn} + L_{arm}\frac{di_{jn}}{dt} = v_j^{ac}, \quad (1)$$

$$\frac{C_{SM}}{N}\frac{dv_{jp}}{dt} = S_{jp}i_{jp}, \quad \frac{C_{SM}}{N}\frac{dv_{jn}}{dt} = S_{jn}i_{jn}, \quad (2)$$

where $S_{jp}$ and $S_{jn}$ are the switching functions of the upper and lower arms, respectively. $v_{jp}$, $v_{jn}$, $i_{jp}$ and $i_{jn}$ are the equivalent capacitor voltages and arm currents of the upper and lower arms, respectively. $v_j^{ac}$ is the ac-side voltage of the MMC.

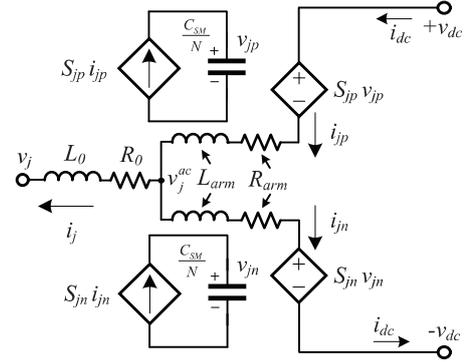

Fig. 2. The equivalent circuit of the ASF model of the MMC for phase $j$ ($j$ = a, b, and c).

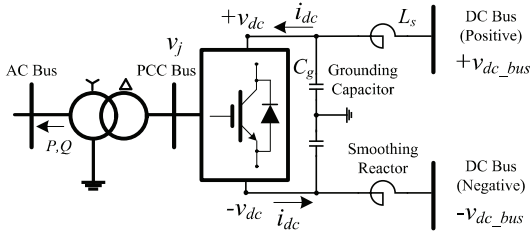

Fig. 1. The overall structure of an MMC station.

The ac-side dynamics is described by

$$v_j^{ac} = v_j + R_0 i_j + L_0 \frac{di_j}{dt}, \quad (3)$$

where $v_j$ is the phase $j$ voltage of PCC.

The dc, fundamental-frequency, and second-order harmonic components dominate the system behavior [16], which are considered in the MMC model. The arm voltages and currents can be expressed as

$$v_{jp} = v^{dc} + v_j^{\omega_0} + v_j^{2\omega_0}, \quad i_{jp} = \frac{1}{3}i_{dc} + \frac{1}{2}i_j + i_j^{2\omega_0},$$
$$v_{jn} = v^{dc} - v_j^{\omega_0} + v_j^{2\omega_0}, \quad i_{jn} = \frac{1}{3}i_{dc} - \frac{1}{2}i_j + i_j^{2\omega_0}, \quad (4)$$

where $v^{dc}$, $v_j^{\omega_0}$, $v_j^{2\omega_0}$ and $i_{dc}$, $i_j$, $i_j^{2\omega_0}$ are the dc, fundamental-frequency, and second-order harmonic voltages and currents, respectively. $S_{jp}$ and $S_{jn}$ are expressed by

$$S_{jp} = \frac{1}{2} + \frac{v_j^{ref}}{\tilde{V}_{dc}}, \qquad S_{jn} = 1 - S_{jp} = \frac{1}{2} - \frac{v_j^{ref}}{\tilde{V}_{dc}}, \quad (5)$$

where $\tilde{V}_{dc}$ is the nominal dc-side voltage of the MMC, and $v_j^{ref}$ is the voltage reference generated by the $dq$ controller.

Substituting (4) and (5) into (1), (2), and (3) and eliminating $v_j^{ac}$, the MMC model in the $dq$ reference frame can be expressed by

$$\frac{d\boldsymbol{x}_1}{dt} = A_{10}\boldsymbol{x}_1 + g\left(\boldsymbol{x}_1\right)\boldsymbol{u}_1, \quad (6)$$

where

$$\boldsymbol{x}_1^T = \left[i_{dc}, i_d, i_q, i_d^{2\omega_0}, i_q^{2\omega_0}, v^{dc}, v_d^{\omega_0}, v_q^{\omega_0}, v_d^{2\omega_0}, v_q^{2\omega_0}\right],$$
$$\boldsymbol{u}_1^T = \left[v_{dc}, v_d^{ref}, v_q^{ref}\right]. \quad (7)$$

B. Modeling of Power and Droop Control

The $dq$ control method is employed for controlling the ac-side active and reactive power of the MMC, as shown in Fig. 3. The PCC bus active and reactive power can be calculated by

$$P = \frac{3}{2}\left(v_d i_d + v_q i_q\right), \quad Q = \frac{3}{2}\left(v_q i_d - v_d i_q\right). \quad (8)$$

The mathematical model of the $dq$ control is expressed as

$$\frac{d\boldsymbol{x}_2}{dt} = A_2\boldsymbol{x}_2 + B_2\boldsymbol{u}_2,$$
$$\boldsymbol{y}_2 = C_2\boldsymbol{x}_2 + D_2\boldsymbol{u}_2, \quad (9)$$

where

$$\boldsymbol{x}_2^T = [\delta i_d, \delta i_q, \delta P, \delta Q], \quad \boldsymbol{u}_2^T = [P_{ref}, Q_{ref}, i_d, i_q],$$
$$\boldsymbol{y}_2^T = \left[v_d^{ref}, v_q^{ref}\right]. \quad (10)$$

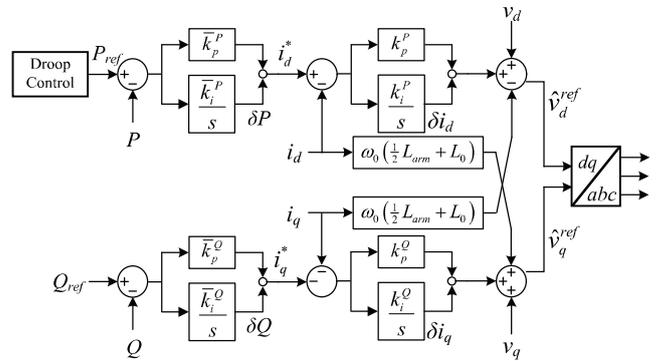

Fig. 3. The $dq$ control of the MMC.

The active power reference in Fig. 3 is generated by the droop control, which can be expressed as [7]

$$P_{ref} = -k\left(v_{dc}^{ref} - v_{dc}\right) + P_0, \quad (11)$$

where $k$ is the slope, and $v_{dc}^{ref}$ and $P_0$ represent two parameters of the droop control.

*C. Modeling of the DC Network*

The structure of a general ac-MTDC system is shown in Fig. 4 and the T equivalent circuit is used for modeling dc transmission line. $R_{Tm}$, $L_{Tm}$ and $C_{Tm}$ are the resistance, inductance and capacitance of the T circuit, where $m$ is the index of the dc transmission line. $v_{Tm}$ is the voltage of the middle point of the T circuit. $i_{\alpha m}$ and $i_{\beta m}$ are the terminal currents. In this paper, a general method is developed for modeling the dc network with arbitrary topology, which is based on the edge-node incidence matrix [15].

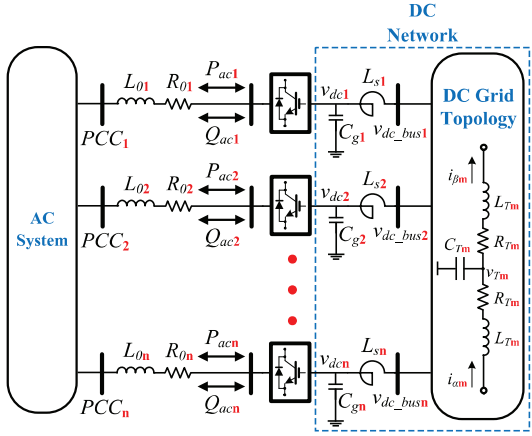

Fig. 4. The structure of a general ac-MTDC system.

For the dc network with $n_{lines}$ transmission lines and $n_{nodes}$ nodes, the size of the edge-node incidence matrix $J$ is $n_{lines} \times n_{nodes}$. The elements of the edge-node incidence matrix $J$ are -1, 0, or 1, which represent the direction of the dc-line current. If the current of $i$th transmission line connecting with the nodes $\zeta$ and $\gamma$ flows from $\zeta$ to $\gamma$, the elements of $J_{(i,\zeta)}$ and $J_{(i,\gamma)}$ should be 1 and $-1$, respectively. If there is no connection between line and node, the associated element is 0.

To conveniently derive the dc network model, two auxiliary matrices $J_1$ and $J_2$ are defined. $J_1$ contains all positive elements of $J$ while $-J_2$ contains all negative elements of $J$. Thus, $J = J_1 - J_2$.

Based on the dc network in Fig. 4, its dynamic model can be expressed by

$$
\begin{aligned}
&L_{\Lambda[s]} \frac{d}{dt}\left(-J_1^T \vec{i_{[\alpha]}} + J_2^T \vec{i_{[\beta]}}\right) = \vec{v_{[dc\_bus]}} - \vec{v_{[dc]}}, \\
&R_{\Lambda[T]} \vec{i_{[\alpha]}} + L_{\Lambda[T]} \frac{d \vec{i_{[\alpha]}}}{dt} = J_1 \vec{v_{[dc\_bus]}} - \vec{v_{[T]}}, \\
&R_{\Lambda[T]} \vec{i_{[\beta]}} + L_{\Lambda[T]} \frac{d \vec{i_{[\beta]}}}{dt} = \vec{v_{[T]}} - J_2 \vec{v_{[dc\_bus]}}, \quad (12)\\
&C_{\Lambda[T]} \frac{d \vec{v_{[T]}}}{dt} = \vec{i_{[\alpha]}} - \vec{i_{[\beta]}}, \\
&C_{\Lambda[g]} \frac{d \vec{v_{[dc]}}}{dt} = -J_1^T \vec{i_{[\alpha]}} + J_2^T \vec{i_{[\beta]}} - \vec{i_{[dc]}},
\end{aligned}
$$

By eliminating $\vec{v_{[dc\_bus]}}$ of (12), the dynamic model of the dc network of Fig. 5 is expressed as

$$\frac{d}{dt} \vec{x_{[3]}} = A_{[3]} \vec{x_{[3]}} + B_{[3]} \vec{u_{[3]}}, \quad (13)$$

where,

$$\vec{x_{[3]}^T} = \left[\vec{i_{[\alpha]}}, \vec{i_{[\beta]}}, \vec{v_{[T]}}, \vec{v_{[dc]}}\right], \quad \vec{u_{[3]}^T} = \left[\vec{i_{[dc]}}\right]. \quad (14)$$

*D. Simulation Verification*

In this subsection, the developed MMC-MTDC model described in (6), (9), (11) and (13) is calculated in MATLAB and verified by comparing with the circuit model simulated in PSCAD/EMTDC.

The IEEE 14-bus ac system is modified to a 14-bus dc system as the study system in this paper and each bus is connected with an MMC station, as shown in Fig. 5. The MMC stations on Buses 1, 2, 3, 6, and 8 operate in the rectifier mode with droop control and the rest operate in the inverter mode with fixed power control.

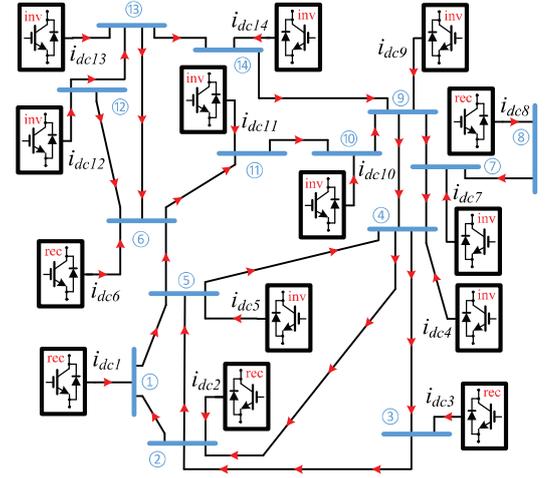

Fig. 5. The 14-bus MMC-MTDC system.

TABLE I
ELECTRIC PARAMETERS OF THE STUDY SYSTEM

| MMC | Value | DC Grid | Value |
|---|---|---|---|
| SM Capacitance [mF] | 20 (Rectifier) | Resistance [$\Omega$/km] | 0.01273 |
| | 25 (Inverter) | Inductance [mH/km] | 0.9337 |
| Arm Inductance [mH] | 0.16 (Rectifier) | Capacitance [$\mu$F/km] | 0.01274 |
| | 0.1 (Inverter) | Smoothing Reactor [mH] | 200 |
| Arm Resistance [$\Omega$] | 2.175 (Rectifier) | Grounding Capacitor [$\mu$F] | 1 |
| | 1.815 (Inverter) | PCC Resistance [$\Omega$] | 1 |
| Voltage Level | 435 (Rectifier) | PCC Inductance [mH] | 0 |
| | 363 (Inverter) | DC Voltage [kV] | $\pm 500$ |

The parameters of the study system are listed in Table I. A step change on the power command of the bus-4 MMC station occurs from 120 MW to 290 MW at $t = 7$ s.

The dynamic responses of the dc voltages and active powers of Buses 1, 4, 6, and 8 are shown in Figs. 6 and 7, respectively. The simulation in PSCAD/EMTDC and calculation in MATLAB are compared to verify the accuracy of the developed model, as depicted by (6), (9), (11) and (13). Based on Figs. 6 and 7, the errors between the simulation and calculation are very small, which validate the accuracy of the developed model.

$$\frac{d}{dt}\begin{bmatrix}\Delta x_{\overrightarrow{[1]}}\\ \Delta x_{\overrightarrow{[2]}}\\ \Delta x_{\overrightarrow{[3]}}\end{bmatrix}=\begin{bmatrix}A_{[1]}+B_{[1]}G_{12}D_{[2]}G_{23}G_{s1} & B_{[1]}G_{12}C_{[2]} & B_{[1]}G_{11}+B_{[1]}G_{12}D_{[2]}G_{21}k_{\Lambda}G_{s2}\\ B_{[2]}G_{23}G_{s1} & A_{[2]} & B_{[2]}G_{21}k_{\Lambda}G_{s2}\\ B_{[3]}G_{3} & 0 & A_{[3]}\end{bmatrix}\begin{bmatrix}\Delta x_{\overrightarrow{[1]}}\\ \Delta x_{\overrightarrow{[2]}}\\ \Delta x_{\overrightarrow{[3]}}\end{bmatrix} \quad (27)$$

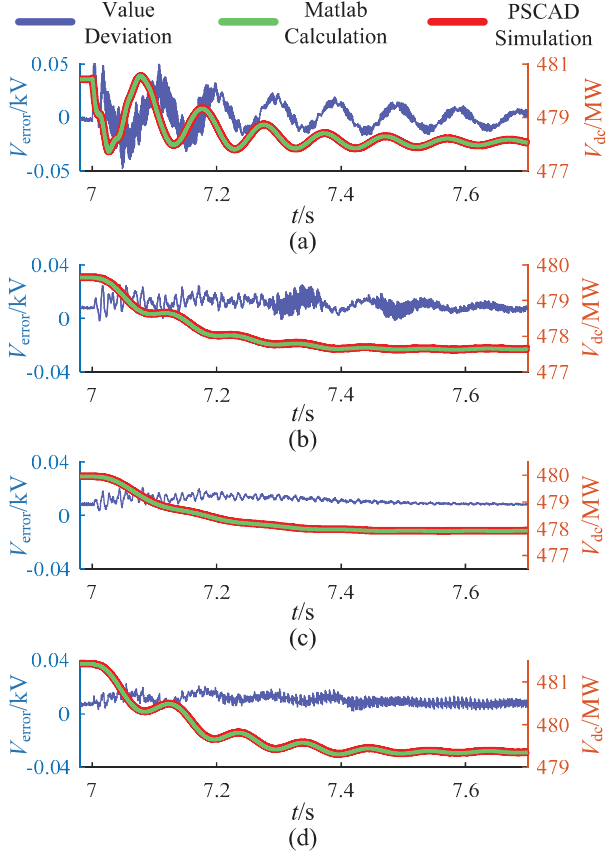

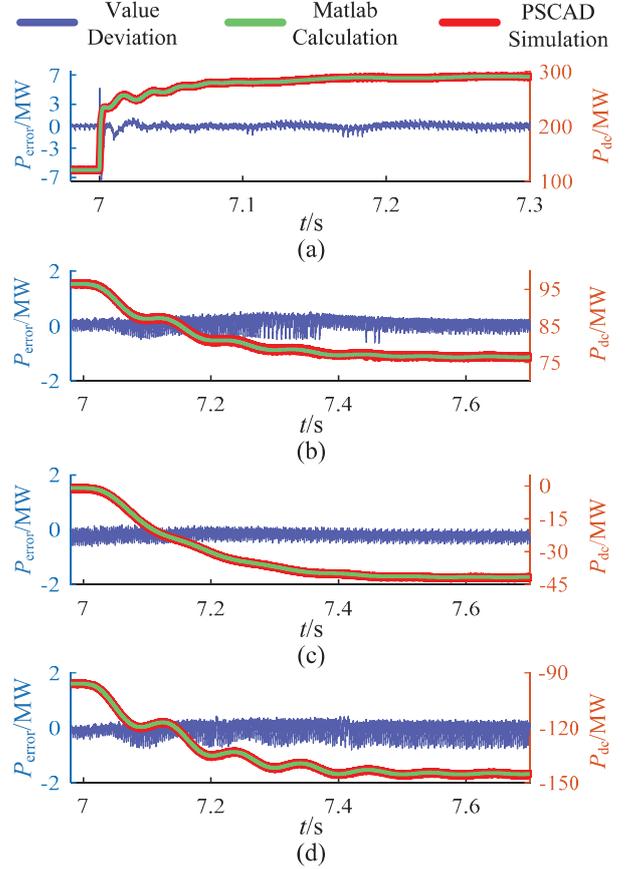

Fig. 6. The dc voltage dynamic responses of some buses and the corresponding numerical comparisons between PSCAD/EMTDC simulation and MATLAB solution. (a) Bus 4, (b) Bus 1, (c) Bus 6, and (d) Bus 8.

Fig. 7. The active power dynamic responses of some buses and the corresponding numerical comparisons between PSCAD/EMTDC simulation and MATLAB solution. (a) Bus 4, (b) Bus 1, (c) Bus 6, and (d) Bus 8.

## III. THE SMALL-SIGNAL MODELING OF MTDC SYSTEM

In this section, the small-signal model expanded at the equilibrium point when $t=7s$ will be established.

By linearizing the nonlinear model of $i$th MMC based on the (6), the small-signal model is derived as

$$\begin{aligned}\frac{d\Delta x_{1(i)}}{dt}&=A_{10(i)}\Delta x_{1(i)}+\frac{\partial\left[g\left(x_{1(i)}\right)u_{1(i)}\right]}{\partial x_{1(i)}}\Delta x_{1(i)}\\ &+\frac{\partial\left[g\left(x_{1(i)}\right)u_{1(i)}\right]}{\partial u_{1(i)}}\Delta u_{1(i)}\\ &=A_{1(i)}\Delta x_{1(i)}+B_{1(i)}\Delta u_{1(i)},\end{aligned} \quad (15)$$

To conveniently establish the small-signal model of multiple sub-systems, by following the notation ways in *Remark* the small-signal model for all MMCs is written as

$$\frac{d}{dt}\Delta x_{\overrightarrow{[1]}}=A_{[1]}\Delta x_{\overrightarrow{[1]}}+B_{[1]}\Delta u_{\overrightarrow{[1]}}, \quad (16)$$

where

$$\begin{aligned}\Delta x_{\overrightarrow{[1]}}^{T}&=\left[\Delta i_{\overrightarrow{[dc]}},\Delta i_{\overrightarrow{[d]}},\Delta i_{\overrightarrow{[q]}},\Delta i_{\overrightarrow{[d]}}^{2\omega_{0}},\Delta i_{\overrightarrow{[q]}}^{2\omega_{0}},\Delta v_{\overrightarrow{[\cdot]}}^{dc},\right.\\ &\left.\Delta v_{\overrightarrow{[d]}}^{\omega_{0}},\Delta v_{\overrightarrow{[q]}}^{\omega_{0}},\Delta v_{\overrightarrow{[d]}}^{2\omega_{0}},\Delta v_{\overrightarrow{[q]}}^{2\omega_{0}}\right],\\ \Delta u_{\overrightarrow{[1]}}^{T}&=\left[\Delta v_{\overrightarrow{[dc]}},\Delta v_{\overrightarrow{[d]}}^{ref},\Delta v_{\overrightarrow{[q]}}^{ref}\right].\end{aligned} \quad (17)$$

The relationship between $A_{1(i)}$ and $A_{[1]}$, $\Delta x_{1(i)}$ and $\Delta x_{\overrightarrow{[1]}}$ can be found in Fig. 8 where the notation $A_{1(i)[m,n]}$ means the element of $m$th row and $n$th column in the system matrix $A_{1}(i)$ of $i$th MMC small-signal model. This structure shown in Fig. 8 makes the matrices $A_{[1]}$ and $B_{[1]}$ easily adjusted through revising contained diagonal sub-matrices when the number of MMCs changes.

By this way and according to (9), (13), the small-signal model of $dq$ controllers for all MMCs and dc grid are directly given by (18) and (20), respectively.

$$\begin{aligned}\frac{d}{dt}\Delta x_{\overrightarrow{[2]}}&=A_{[2]}\Delta x_{\overrightarrow{[2]}}+B_{[2]}\Delta u_{\overrightarrow{[2]}},\\ \Delta y_{\overrightarrow{[2]}}&=C_{[2]}\Delta x_{\overrightarrow{[2]}}+D_{[2]}\Delta u_{\overrightarrow{[2]}}.\end{aligned} \quad (18)$$

where

$$\begin{aligned}\Delta x_{\overrightarrow{[2]}}^{T}&=\left[\Delta\delta i_{\overrightarrow{[d]}},\Delta\delta i_{\overrightarrow{[q]}},\Delta\delta P_{\overrightarrow{[\cdot]}},\Delta\delta Q_{\overrightarrow{[\cdot]}}\right],\\ \Delta y_{\overrightarrow{[2]}}^{T}&=\left[\Delta v_{\overrightarrow{[d]}}^{ref},\Delta v_{\overrightarrow{[q]}}^{ref}\right],\\ \Delta u_{\overrightarrow{[2]}}^{T}&=\left[\Delta P_{\overrightarrow{[ref]}},\Delta Q_{\overrightarrow{[ref]}},\Delta i_{\overrightarrow{[d]}},\Delta i_{\overrightarrow{[q]}}\right].\end{aligned} \quad (19)$$

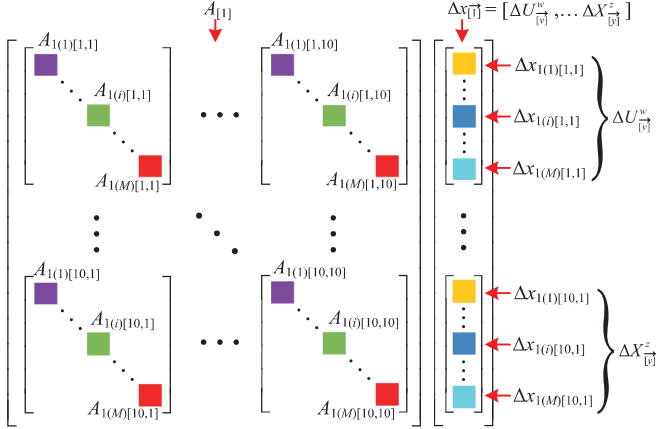

Fig. 8. The relationship of small-signal model between single one (15) and multiple ones (16).

and
$$\frac{d}{dt}\Delta \boldsymbol{x}_{\overrightarrow{[3]}} = \boldsymbol{A}_{[3]}\Delta \boldsymbol{x}_{\overrightarrow{[3]}} + \boldsymbol{B}_{[3]}\Delta \boldsymbol{u}_{\overrightarrow{[3]}}, \quad (20)$$

where
$$\Delta \boldsymbol{x}_{[3]}^T = \left[\Delta \boldsymbol{i}_{\overrightarrow{[\alpha]}}, \Delta \boldsymbol{i}_{\overrightarrow{[\beta]}}, \Delta \boldsymbol{v}_{\overrightarrow{[T]}}, \Delta \boldsymbol{v}_{\overrightarrow{[dc]}}\right], \Delta \boldsymbol{u}_{[3]}^T = \left[\Delta \boldsymbol{i}_{\overrightarrow{[dc]}}\right]. \quad (21)$$

The small-signal model of droop controller is given by
$$\Delta \boldsymbol{P}_{\overrightarrow{[ref]}} = \boldsymbol{k}_\Lambda \Delta \boldsymbol{v}_{\overrightarrow{[dc]}}, \quad (22)$$

where $\boldsymbol{k}_\Lambda$ indicates the diagonal matrix consisted of all $k$ of the droop controls in MTDC system. If some MMCs do not have droop controls, then the corresponding diagonal positions of $\boldsymbol{k}_\Lambda$ will be 0.

Noting that (16), (18) and (20) have couplings with each other by their inputs and state variables, therefore, the autonomous system will be derived for establishing the small-signal model of whole MTDC system. Firstly, the relationship between $\Delta \boldsymbol{u}_{\overrightarrow{[3]}}$ and $\Delta \boldsymbol{x}_{\overrightarrow{[1]}}$ is shown as

$$\Delta \boldsymbol{u}_{\overrightarrow{[3]}} = \Delta \boldsymbol{i}_{\overrightarrow{[dc]}} = [\boldsymbol{I}, \underbrace{\boldsymbol{0},\boldsymbol{0},\cdots,\boldsymbol{0}}_{n_{nodes}\times 9 n_{nodes}}]\Delta \boldsymbol{x}_{\overrightarrow{[1]}} = \boldsymbol{G}_3 \Delta \boldsymbol{x}_{\overrightarrow{[1]}}, \quad (23)$$

where $\boldsymbol{0}$ and $\boldsymbol{I}$ are the zero and identical matrix, respectively.

Moreover, there are the relationships shown as
$$\begin{bmatrix}\Delta \boldsymbol{i}_{\overrightarrow{[d]}}\\ \Delta \boldsymbol{i}_{\overrightarrow{[q]}}\end{bmatrix} = \underbrace{\begin{bmatrix}\boldsymbol{0},\boldsymbol{I},\boldsymbol{0},\cdots,\boldsymbol{0}\\ \boldsymbol{0},\boldsymbol{0},\boldsymbol{I},\cdots,\boldsymbol{0}\end{bmatrix}}_{2n_{nodes}\times 10 n_{nodes}} \Delta \boldsymbol{x}_{\overrightarrow{[1]}} = \boldsymbol{G}_{s1}\boldsymbol{x}_{\overrightarrow{[1]}}, \quad (24)$$
$$\Delta \boldsymbol{v}_{\overrightarrow{[dc]}} = [\boldsymbol{0},\boldsymbol{0},\boldsymbol{0},\boldsymbol{I}]\Delta \boldsymbol{x}_{\overrightarrow{[3]}} = \boldsymbol{G}_{s2}\Delta \boldsymbol{x}_{\overrightarrow{[3]}}.$$

Then, involving (22) and (24) into the expression of $\Delta \boldsymbol{u}_{\overrightarrow{[2]}}$, $\Delta \boldsymbol{u}_{\overrightarrow{[2]}}$ will be shown as

$$\Delta \boldsymbol{u}_{\overrightarrow{[2]}} = \begin{bmatrix}\boldsymbol{I}\\ \boldsymbol{0}\\ \boldsymbol{0}\\ \boldsymbol{0}\end{bmatrix}\Delta \boldsymbol{P}_{\overrightarrow{[ref]}} + \begin{bmatrix}\boldsymbol{0}\\ \boldsymbol{I}\\ \boldsymbol{0}\\ \boldsymbol{0}\end{bmatrix}\Delta \boldsymbol{Q}_{\overrightarrow{[ref]}} + \begin{bmatrix}\boldsymbol{0} & \boldsymbol{0}\\ \boldsymbol{0} & \boldsymbol{0}\\ \boldsymbol{I} & \boldsymbol{0}\\ \boldsymbol{0} & \boldsymbol{I}\end{bmatrix}\begin{bmatrix}\Delta \boldsymbol{i}_{\overrightarrow{[d]}}\\ \Delta \boldsymbol{i}_{\overrightarrow{[q]}}\end{bmatrix}, \quad (25)$$
$$= \boldsymbol{G}_{23}\boldsymbol{G}_{s1}\Delta \boldsymbol{x}_{\overrightarrow{[1]}} + \boldsymbol{G}_{21}\boldsymbol{k}_\Lambda \boldsymbol{G}_{s2}\boldsymbol{x}_{\overrightarrow{[3]}}$$

Considering that the reactive power references $Q_{ref}$ keep constant, $\Delta \boldsymbol{Q}_{\overrightarrow{[ref]}}$ is a zero vector. In addition, introduce the results shown in (18), (24) and (25) into $\Delta \boldsymbol{u}_{\overrightarrow{[1]}}$ and it has

$$\Delta \boldsymbol{u}_{\overrightarrow{[1]}} = \begin{bmatrix}\boldsymbol{I}\\ \boldsymbol{0}\\ \boldsymbol{0}\end{bmatrix}\Delta \boldsymbol{v}_{\overrightarrow{[dc]}} + \begin{bmatrix}\boldsymbol{0} & \boldsymbol{0}\\ \boldsymbol{I} & \boldsymbol{0}\\ \boldsymbol{0} & \boldsymbol{I}\end{bmatrix}\begin{bmatrix}\Delta \boldsymbol{v}_{\overrightarrow{[d]}}^{ref}\\ \Delta \boldsymbol{v}_{\overrightarrow{[q]}}^{ref}\end{bmatrix}$$
$$= \boldsymbol{G}_{12}\boldsymbol{D}_{\Lambda[2]}\boldsymbol{G}_{23}\boldsymbol{G}_{s1}\Delta \boldsymbol{x}_{\overrightarrow{[1]}} + \boldsymbol{G}_{12}\boldsymbol{C}_{\Lambda[2]}\Delta \boldsymbol{x}_{\overrightarrow{[2]}} + \quad (26)$$
$$\left(\boldsymbol{G}_{11} + \boldsymbol{G}_{12}\boldsymbol{D}_{\Lambda[2]}\boldsymbol{G}_{21}\boldsymbol{k}_\Lambda \boldsymbol{G}_{s2}\right)\Delta \boldsymbol{x}_{\overrightarrow{[3]}}$$

At last, involving all the expressions shown in (23), (25) and (26) into (16), (18) and (20), the final small-signal model of whole MTDC system is given by (27).

## IV. DERIVATION FOR SLOPE CONSTRAINTS ESTABLISHMENT OF DROOP CONTROL

It is indispensable to emphasize that this paper concentrates on the impact of the slopes $k$ of multiple droop controllers in MTDC system on the small-disturbance stability of the equilibrium point. So, in the subsequent sections, to estimate the stability regions of droop controller slopes, the values of $k_\Lambda$ will be varied. Meanwhile, the parameters $P_0$ shown in (11) will be correspondingly changed for guaranteeing the same equilibrium point and, according to (18), they are not appeared in (27) to affect the proposed small-signal model.

### A. Instability Caused by Droop Control

The eigenvalue loci of small-signal model (27) are obtained when $k_1$ increase from 10 to 2010. According to the loci, there are 6 pairs of conjugate eigenvalue in total 270 eigenvalues passing through the imaginary-axis and moving to the right-half-plane, the trajectories of which are expressed in Fig. 9.

These eigenvalue loci, at least, show that when the slope $k$ of droop control exceeds some stability region, it will cause instability. In light of the couplings and the interactions of multiple droop controllers, the stability region of multiple droop control slopes at the specified equilibrium point will be abstract to describe and difficult to demonstrate.

### B. Method of Eigenvalues Sensitivity Calculation

For convenient statement, (27) is rewritten as a multivariate autonomous system given by (28).

$$\frac{d}{dt}\boldsymbol{x}_0 = \boldsymbol{A}_{ss}\left(k_1 \cdots k_{n_{droop}}\right)\boldsymbol{x}_0 \quad (28)$$

where $\boldsymbol{x}_0^T = [\boldsymbol{x}_{\overrightarrow{[1]}}, \boldsymbol{x}_{\overrightarrow{[2]}}, \boldsymbol{x}_{\overrightarrow{[3]}}]$, $k_1$ to $k_{n_{droop}}$ are the initial slopes of each droop control and $n_{droop}$ is the number of all droop controls.

Thus, based on the Taylor series of multivariate function, the deviation of $i$th eigenvalue $\lambda_i$ can be given by (29).

$$\Delta \lambda_i = \sum_{j=1}^{n_{droop}}\frac{\partial \lambda_i}{\partial k_j}\Delta k_j + \frac{1}{2!}\sum_{j,l=1}^{n_{droop}}\frac{\partial^2 \lambda_i}{\partial k_j \partial k_l}\Delta k_j \Delta k_l \quad (29)$$

Moreover, the calculation ways of first and second-order partial derivative with respect to $k$, according to reference [17], are directly given by

$$\frac{\partial \lambda_i}{\partial k_j} = z_i^T \frac{\partial \boldsymbol{A}_{ss}}{\partial k_j}w_i,$$
$$\frac{\partial^2 \lambda_i}{\partial k_j \partial k_l} = z_i^T \frac{\partial^2 \boldsymbol{A}_{ss}}{\partial k_j \partial k_l}w_i + z_i^T \frac{\partial \boldsymbol{A}_{ss}}{\partial k_l}\frac{\partial w_i}{\partial k_j} + z_i^T \frac{\partial \boldsymbol{A}_{ss}}{\partial k_j}\frac{\partial w_i}{\partial k_l}. \quad (30)$$

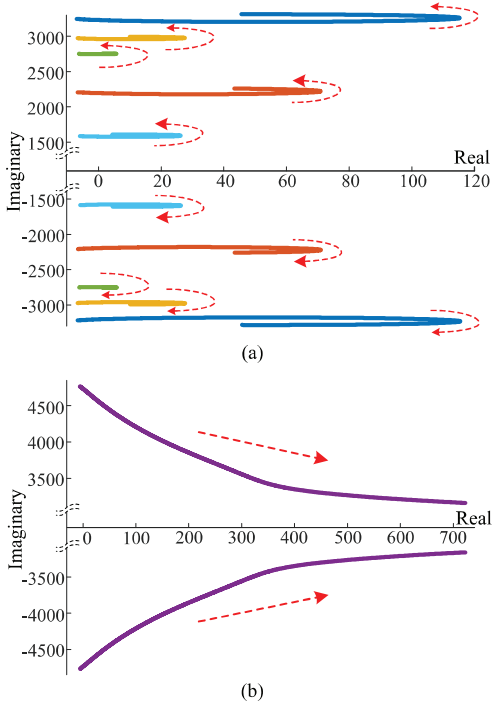

Fig. 9. The eigenvalue loci of unstable eigenvalues along with the increment $k_1$ from 10 to 2010.

where $A_{ss}$ is the system matrix of (28), $z_i^T$ and $w_i$ are the corresponding left and right eigenvector of $i$th eigenvalue which are normalized ($z_i^T w_i = 1$) and orthogonal ($z_i^T w_j = 0, i \neq j$). The calculation of the partial derivative of right eigenvectors is exhibited as

$$\frac{\partial w_i}{\partial k_j} = \sum_{\kappa=1, \neq i} \frac{1}{\lambda_i - \lambda_\kappa} \left( z_\kappa^T \frac{\partial A_{ss}}{\partial k_j} w_i \right) w_\kappa, \qquad (31)$$

### C. Calculation for Slope Constraints

The principle of slope constraints based on eigenvalue deviation and sensitivity is shown in Fig. 10. For convenient understanding, the eigenvalue loci are simplified as strait lines. Even though the initial eigenvalue may locate at the right- or left-half-plane such as Fig. 10 (a) and (b), respectively, the principle of proposed method is suitable for both situation. In Fig. 10 (a), the signed distance from the position of $\lambda_{21} < 0$ to the imaginary-axis is regarded as the stability margin $\delta_{21} > 0$ which has the same value by using another conjugate eigenvalue $\lambda_{22}$. If the eigenvalue $\lambda_{22}$ has a deviation $\Delta \lambda_1$ caused by the variations of $k_1 \cdots k_{n_{droop}}$, and the real part of which holds $Re[\Delta \lambda_2] < \delta_{21}$, the MTDC system will still be ensured stable. Otherwise, if the variations result a eigenvalue deviation $\Delta \lambda_2$ which brings out $Re[\Delta \lambda_1] > \delta_{21}$, it will lead to the instability. In Fig. 10 (b), the principle is same with the demonstration in Fig. 10 (a) but the only differences are the signs of $\delta_{21}$, $Re[\Delta \lambda_1]$ and $Re[\Delta \lambda_2]$ which are minus.

According to Fig. 10 and confronting a comprehensive impact of the slope $k_1 \cdots k_{n_{droop}}$ varied, every eigenvalue deviation on the real-axis $Re[\Delta \lambda_i]$ must be less than its corresponding stability margin, which is given by

$$Re[\Delta \lambda_i] < \delta_i = 0 - Re[\lambda_i]. \qquad (32)$$

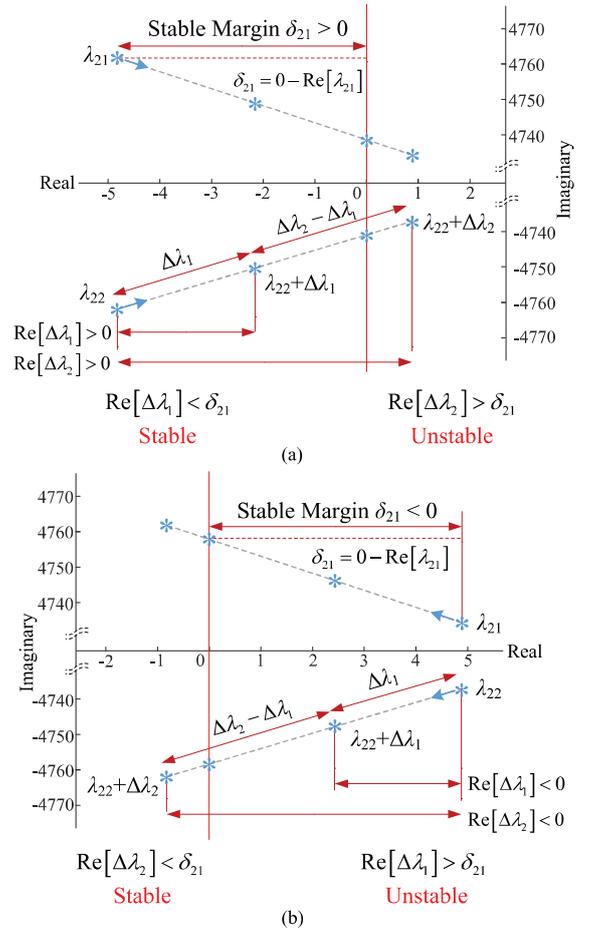

Fig. 10. The principle of the slope constraints based on eigenvalue deviation. (a) The initial eigenvalues locate at left-half-plane. (b) The initial eigenvalues locate at right-half-plane.

(29) will be applied to estimate the eigenvalue deviation on real-axis. Thus, the deviation of $i$th eigenvalue on the real-axis is given by

$$Re[\Delta \lambda_i] = \sum_{j=1}^{n_{droop}} Re\left[\frac{\partial \lambda_i}{\partial k_j}\right] \Delta k_j + \frac{1}{2!} \sum_{j,l=1}^{n_{droop}} Re\left[\frac{\partial^2 \lambda_i}{\partial k_j \partial k_l}\right] \Delta k_j \Delta k_l \qquad (33)$$

As a result, the desired slope constraints of multiple droop controls for the slope stability region are obtained and shown as

$$\sum_{j=1}^{n_{droop}} Re\left[\frac{\partial \lambda_i}{\partial k_j}\right] \Delta k_j + \frac{1}{2!} \sum_{j,l=1}^{n_{droop}} Re\left[\frac{\partial^2 \lambda_i}{\partial k_j \partial k_l}\right] \Delta k_j \Delta k_l < \delta_i \qquad (34)$$

and due to the conjugate eigenvalues, the number of inequalities will be much less that the order of (27).

The supremum $k_i^{sup}$ of $k_i$ can be acquired by adding the minimum slope deviation $\Delta k_i$ calculated by constraints shown in (34) and initial value $k_i$, which is shown as

$$k_i^{sup} = k_i + min\{\Delta k_i | \text{Slope Constraints} \quad (34)\} \qquad (35)$$

When calculating (30), the initial values of $k$ from $k_1$ to $k_{n_{droop}}$ will naturally affect the results of (32) and (34). However, based on the Taylor Series, the expansion point does not influence on the final result. Therefore, whatever the coefficients of (34) are, their corresponding stability regions of droop controller slopes will be the same. Fig. 11 shows the flowchart of proposed method.

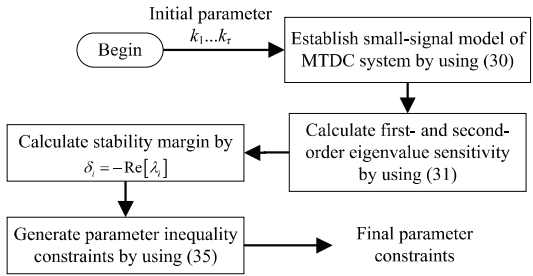

Fig. 11. The flowchart of the stability region estimation for droop control slopes.

It is evident that the proposed method will obtain a group of inequalities. Obviously, the significant advantage of proposed method is avoiding repeated eigenvalue-solving for locus. Furthermore, the analytic solutions of the quadratic inequalities are quite efficient and quick to estimate the stability regions of the multiple droop controller slopes.

## V. STUDY CASES

Two study cases with different coefficient of the droop controllers are shown as *Case* 1 and *Case* 2, respectively. Both study cases are established on the IEEE 14-bus aforementioned as Fig. 5. Then, to confirm the feasibility of proposed method, the slope constraints of two study cases will estimate the supremum of $k_i, i = 1, 2, 3, 6, 8$. For comparison purpose, the exact values of each supremum calculated through the eigenvalue loci will be expressed. Meanwhile, the self- and cross-validations for both cases are listed as well to verify the performance of both slope constraints.

Tab. II and Tab. III provide the examination details of two cases. Self-validation is that using the constraints derived from one case to estimate the supremum of $k$ in the same case. Cross-validation is that employing the constraints derived from one case to calculate the supremum of $k$ in the other case. To get the supremum of $k_1$ through a group of specified slope constraints, for example, the initialized deviation will be determined at first that between the rest of $k$ in the study case and the ones in calculation case which generates the specified constraints. By applying (34) and (35), the desired supremum can be estimated.

The performance of the proposed method shown in Tab. II and Tab. III are satisfactory. The self- and cross- validations of both cases indicate that the different groups of the slope constraints have the same capability of estimation. Compared with the performance of $k_3$, $k_6$ and $k_8$, the supremes of $k_1$ and $k_2$ calculated by the proposed method have some small deviations with the ones determined by the eigenvalue loci. The reason should be the reminder of Taylor Series.

In addition, the stability regions of $k_1$ and $k_2$ in *Case* 1 obtained through the proposed slope constraints and the eigenvalue loci are displayed in Fig. 12. The orange hatched area is the region estimated by the slope constraints. The blue hatched area is the region calculated by the eigenvalue loci. The initial value of the utilized constraints is $k_i = 50, i = 1, 2, 3, 6, 8$. It is clear to confirm that the accuracy of the proposed method is feasible. On the other side, Fig. 12 also confirm that the residual caused by the Taylor Series reminder occurs far away from the expansion point.

TABLE II
THE FEASIBILITY EXAMINATION OF *Case* 1

| Case 1 | | | | | |
|---|---|---|---|---|---|
| **Initial Value** | $k_1 = 10$, | $k_2 = 15$, | $k_3 = 18$, | $k_6 = 20$, | $k_8 = 24$ |
| **Calculated Supremum** | $k_1^{sup}$ | $k_2^{sup}$ | $k_3^{sup}$ | $k_6^{sup}$ | $k_8^{sup}$ |
| **Exact Supremum** | 78.0 | 81.9 | 68.1 | 67.8 | 71.6 |
| **Purpose** | Self-Validation by *Case* 1's inequalities | | | | |
| **Initialized Deviation** | $\Delta k_1 = \Delta k_2 = \Delta k_3 = \Delta k_6 = \Delta k_8 = 0$ | | | | |
| **Estimated Supremum** | 73.5492 | 77.8330 | 68.0443 | 67.8155 | 71.5543 |
| **Purpose** | Cross-Validation by *Case* 2's inequalities | | | | |
| **Initialized Deviation** | $\Delta k_1 = \Delta k_2 = -10, \Delta k_3 = 7, \Delta k_6 = 10, \Delta k_8 = 6$ | | | | |
| **Estimated Supremum** | 73.5485 | 77.8483 | 68.0561 | 67.6931 | 71.5601 |

TABLE III
THE FEASIBILITY EXAMINATION OF *Case* 2

| Case 2 | | | | | |
|---|---|---|---|---|---|
| **Initial Value** | $k_1 = 20$, | $k_2 = 25$, | $k_3 = 11$, | $k_6 = 10$, | $k_8 = 18$ |
| **Calculated Supremum** | $k_1^{sup}$ | $k_2^{sup}$ | $k_3^{sup}$ | $k_6^{sup}$ | $k_8^{sup}$ |
| **Exact Supremum** | 77.7 | 81.8 | 68.1 | 67.9 | 71.6 |
| **Purpose** | Self-Validation by *Case* 2's inequalities | | | | |
| **Initialized Deviation** | $\Delta k_1 = \Delta k_2 = \Delta k_3 = \Delta k_6 = \Delta k_8 = 0$ | | | | |
| **Estimated Supremum** | 74.8981 | 79.2868 | 68.0561 | 67.7036 | 71.5601 |
| **Purpose** | Cross-Validation by *Case* 1's inequalities | | | | |
| **Initialized Deviation** | $\Delta k_1 = \Delta k_2 = 10, \Delta k_3 = -7, \Delta k_6 = -10, \Delta k_8 = -6$ | | | | |
| **Estimated Supremum** | 74.8741 | 79.2378 | 68.0442 | 67.7826 | 71.5543 |

At last, a PSCAD/EMTDC simulation with $k_2$ step change based on *Case* 1 is exhibited. Fig. 13 (a) shows the situation of $k_2$ step change from 15 to 100 under $k_1 = 10$. The corresponding supreme of $k_2$ can be found in Tab. II. To ensure the constant equilibrium point, the parameter $P_0$ has a step change as well and the corresponding value of which is shown in Fig. 13 (a). Fig. 13 (b) and (c) express the waves of droop controller output and measured active power of MMC tied with Bus 2 after $k_2$ step change. The unstable phenomenon is obvious.

## VI. CONCLUSION

This paper presents the stability analysis of MMC-based MTDC system established by IEEE 14-bus system, and further develops the inequity constraints of multiple droop controller slopes based on the eigenvalue sensitivity. The modeling of MMC-based MTDC system is evaluated and its validation is also proved by simulation of PSCAD/EMTDC and calculation of MATLAB.

To obtain the stability region for droop controller slopes, this paper proposes a method by utilizing the eigenvalue sensitivity to construct a group of inequality constraints related to the eigenvalues and slope deviations. Moreover, two cases

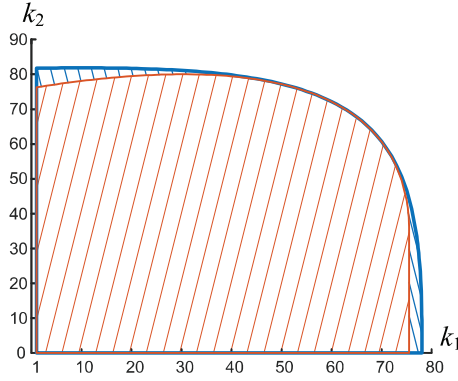

Fig. 12. The stability region of $k_1$ and $k_2$ acquired through the proposed slope constraints and the eigenvalue loci.

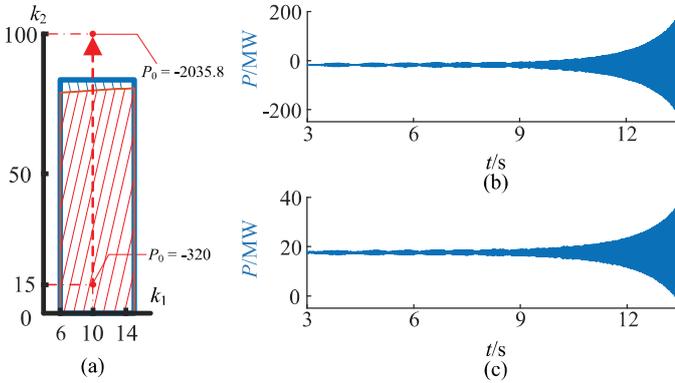

Fig. 13. The unstable active power and corresponding reference of droop controller of MMC tied with Bus 2 when $k_2$ has a step change. (a) The step change of $k_2$ on the stability region. (b) The reference of droop controller. (c) The active power of MMC

are applied to attest the feasibility through searching the supremum of $k$. The self- and cross-validations show the generality and feasibility of the proposed method. At last, the slope stability regions drew by the proposed method and the eigenvalue calculations, respectively, exhibit the precision of the proposed method.

The proposed method just needs algebraic operations depending on eigenvalue and eigenvectors, which avoids so much computational burden and acts more efficient.